\documentclass[12pt, a4paper]{article}
\usepackage[USenglish]{babel}
\usepackage{float,color}
\usepackage[margin=1.0in]{geometry}
\usepackage{authblk}
\usepackage{graphicx}

\begin{document}

\begin{center}
{\Large \bf {Resolved Photon Processes: \\ A Tribute to Prof. Rohini M.
Godbole}}

\vspace*{3mm}
{\bf {\large Manuel Drees}}\footnote{mandrees@uni-bonn.de}\\
\vspace*{3mm}
{\it Physics Institute \& BCTP, Bonn University,
  Nussallee 12, 53115 Bonn, Germany}
\end{center}	
\begin{abstract}
  Hadrons are particles composed of elementary quarks and gluons,
  which have strong interactions described by Quantum Chromodynamics
  (QCD); examples are protons and neutrons, which bind to form atomic
  nuclei. In contrast, photons are usually thought of as the
  elementary force carriers of quantum electrodynamics (QED). However,
  at the quantum level a photon can fluctuate into a quark antiquark
  pair.  At sufficiently high energies these virtual quarks can become
  real, physical particles by interacting with other particles, in
  particular with other hadrons. In this way photons with energies
  exceeding a few GeV acquire properties of a hadron. Resolved photon
  processes are reactions that probe these hadronic properties. These
  processes often dominate the production of hadronic final states,
  including jets (sprays of collimated hadrons), at electron--proton
  and electron--po\-si\-tron colliders. Implications of this for
  backgrounds at future high--energy lepton colliders remain poorly
  understood.
\end{abstract}

\section{Introduction: What Are Resolved Photons?}

Elementary particles can be categorized according to different
criteria.  For example, one can distinguish particles which have
strong interactions, associated with the gauge group $SU(3)_C$, from
those which don't.  Quarks and gluons belong to the first category,
the electron, neutrinos and the photon to the latter. Alternatively,
one can distinguish between matter particles carrying half a unit of
spin and force carriers carrying a whole unit of spin. Quarks, the
electron and neutrinos are in the first category, the gluons, the
$W^\pm$ bosons associated with the weak force, and the photon in the
latter.

However, these distinctions are clear--cut only for the idealized
entities from which quantum field theories (QFT) are constructed. Of
particular interest for this article is the theory of strong
interactions, denoted by Quantum Chromodynamics (QCD), as developed by
Gell--Mann and others in the 1960's and early 1970's \cite{GM1,
  Fritzsch}. Its basic building blocks are the quarks and gluons
mentioned in the previous paragraph. An important feature of QCD is
``asymptotic freedom'': QCD interactions become weaker when probed at
very short distance or, equivalently due to the uncertainty principle,
with very large momentum transfer \cite{GW, Politzer}. This allows us
to perform perturbative calculations in QCD -- but only for reactions
involving a sufficiently large momentum transfer; this typically
requires center--of--mass energies of at least several GeV. At small
momentum transfer, or large distance,\footnote{Note that a femtometer,
  $10^{-15}$ m, is a ``large'' distance in this context!}  strong
interactions become very strong indeed. Although the connection has
not yet been proven to everybody's satisfaction, this is presumably
related to the phenomenon of ``confinement'': quarks and gluons can
only exist within bound states, i.e. in hadrons like the proton or
pion; they do not exist as free particles, unlike the electron or
photon which can travel over macroscopic distances.

There is an additional complication, alluded to in the first sentence
of the previous paragraph. As already noted, quarks and gluons can be
treated as individual particles only if probed with a large momentum
transfer. This typically also implies large energy transfer, which in
turn allows the probed particle to emit other quanta. In particular,
both quarks and gluons can emit an additional gluon when probed at
sufficiently short distance; a gluon can also split into a
quark--antiquark pair. The rules of QCD imply that the probability for
such a splitting to occur becomes very large at large momentum transfer,
provided the emitted gluon (or the quark--antiquark pair) is nearly
collinear with the original particle.

\begin{figure}[t]
   \centering
   \includegraphics*[width=14cm]{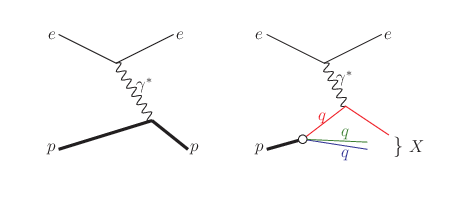}
   \caption{\textit{Electron proton scattering proceeds via the
       exchange of a virtual photon $\gamma^*$. At low momentum
       transfer, which means low virtuality $Q$ of the exchanged
       photon, the dominant contributing process is elastic scattering
       (left diagram). At larger $Q$ the dominant process is instead
       deep inelastic scattering (DIS) (right), where the virtual
       photon interacts with a quark in the proton; the struck quark
       and the proton remnant together form a hadronic final state
       $X$, which typically contains many hadrons. Here the proton has
       been depicted as consisting of three ``colorful'' quarks (i.e. quarks
     charged under $SU(3)_C$).}}
%   \vspace*{-2mm}
   \label{fig1}
\end{figure}

As a result, the answer to the question ``What is inside a proton''
depends strongly on how closely you look at it. In particle physics,
one of the cleanest ways to ``look at'' a proton is to scatter an
electron off it; see Fig.~\ref{fig1}. If the momentum $Q$ exchanged
between the proton and the electron\footnote{Let $k$ and $k'$ be the
  $4-$momenta of the incoming and outgoing electron in
  Fig.~\ref{fig1}; then $Q = \sqrt{-(k-k')^2}$, using the convention
  where a space--like $4-$vector has negative norm.} is less than
$100 \; {\rm MeV}$ or so, the proton looks essentially like a point
particle; the resolution is not sufficient to look ``into'' the
proton. (I use ``natural units'' where the speed of light $c=1$.)
Conversely, for $Q$ well above $1 \; {\rm GeV}$ the electron
predominantly scatters off an individual quark in the proton, rather
than the proton as a whole. This quark is then ``kicked out'' of the
proton. In the process it typically radiates off several almost
collinear gluons. This is somewhat analogous to the electromagnetic
radiation emitted by an accelerated charge even in classical
electrodynamics.

However, we just saw that quarks and gluons cannot exist as free
particles over distances exceeding a (fraction of) a femtometer.  Let
us ignore the gluons emitted by the struck quark for a moment.  This
quark as well as the proton remnant carry $SU(3)_c$ ``color''
charge. The associated potential is similar to a Coulomb potential at
short distances, but increases approximately linearly at large
distance. At some point it therefore becomes energetically favorable
to produce a quark--antiquark pair out of the vacuum, which shields
these color charges. This leads to hadronization. In the simplest
case, the struck quark combines with the antiquark from the vacuum to
form a meson, while the quark from the vacuum combines with the proton
remnant to form a baryon. This picture becomes considerably more
complicated once the gluons emitted by the struck quark are taken into
account. They also hadronize. Therefore for momentum exchange well
above a GeV electron--proton scattering typically leads to a
multi--hadron final state. This process is called ``deep inelastic
scattering'' (DIS).  Crucially, it can be shown that the total
transition amplitude for this process {\em factorizes} into a ``hard
scattering'' piece describing electron--quark scattering, which can be
computed in QCD, and the hadronization piece which cannot be described
perturbatively. As a result, only the {\em inclusive} cross section
can be computed perturbatively: after the electron--quark scattering
the struck quark as well as the proton remnant have to hadronize into
some sort of multi--hadron final state, but we cannot compute reliably
exactly how this final state will look like, e.g. how many hadrons it
will contain. Hence we can predict the cross section only after
summing over all these hadronic final states $X$, although the cross
section can still be differential in the kinematics of the scattered
electron. In other words, only the sum
$\sum_X {\rm d}\sigma(ep \rightarrow eX)$ can be computed
perturbatively.

At even larger momentum exchange, above $5 \; \rm{GeV}$ or so, some
statements about the hadronic final state do become possible. In this
case the struck quark and the proton remnant are sufficiently far away
in phase space that their hadronization products mostly fall into two
groups of hadrons, one moving in the direction of the original proton,
the other in the direction of the struck quark. These systems of
collimated hadrons are called ``jets'' \cite{Feynman}. The simplest
final state emerging from DIS at large momentum transfer therefore
contains two jets. Final states with more jets are also possible,
e.g. when a gluon is emitted that is {\em not} collinear; its
hadronization products therefore populate a new part of phase
space. Emitting such a gluon requires {\em another} large momentum
transfer, i.e. the cross section for such a reaction is calculable
perturbatively; it is suppressed by an extra factor of the strong
coupling constant $\alpha_S$ -- the Sommerfeld constant of QCD, if you
will.

\begin{figure}[t]
   \centering
   \includegraphics*[width=14cm]{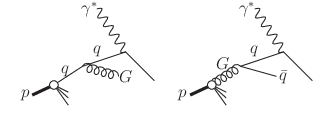}
   \caption{\textit{At sufficiently large momentum transfer $Q$ the
       struck quark is likely to emit a nearly collinear gluon $G$
       before interacting with the virtual photon (left diagram);
       since the gluon carries some energy, the struck quark has less
       energy than the quark originally taken from the
       proton. Similarly, a gluon from the proton can split into a
       collinear quark--antiquark pair, and a member of the pair can
       then interact with the virtual photon (right diagram). These
       processes are described by the famous DGLAP equations. (The
       electron emitting the virtual photon has been omitted here for
       simplicity.)}}
%   \vspace*{-2mm}
   \label{fig2}
\end{figure}

Increasing the momentum exchange $Q$ not only helps to separate the
struck quark's hadronization products from those of the proton
remnant. It also allows the quark to emit physical gluons {\em before}
it is struck by the electron. The probability for this is again large
only for collinear gluons -- in this case, collinear with the incoming
quark, i.e. with the incoming proton; see Fig.~\ref{fig2}. In this
sense the constituents of a proton that is being probed with a large
momentum exchange $Q$ can be considered to be a large number of
quarks, antiquarks and gluons all traveling in the same direction.
The fluxes of these ``partons'' are described by parton distribution
functions (PDFs), which depend on two variables: the fraction $x$ of
the proton's energy carried by this parton, also called the Bjorken
variable \cite{Bj}; and the momentum scale $Q$ at which the PDFs are
probed.  The $x-$dependence cannot be computed perturbatively, it has
to be taken from experiment or estimated using non--perturbative
methods. In contrast, the $Q-$dependence can be computed
perturbatively using the famous DGLAP (for
Dokshitzer--Gribov--Lipatov--Altarelli--Parisi \cite{GL, Dok, AP})
equations.

So far we have been discussing deep inelastic electron--proton
scattering. While Rohini has done some work in this area, in
particular for protons that are bound in heavier nuclei \cite{RN1,
  RN2}, the topic of this article are resolved photons, not resolving
protons. In the context of deep--inelastic scattering, we have to
replace the target proton by a photon, leading to the reaction
$e\gamma \rightarrow eX$, where $X$ again stands for a multi--hadron
final state; as before, the corresponding cross section can be
computed perturbatively only after summing over all such final
states. Since the photon itself doesn't carry any electromagnetic (or
other) charge\footnote{This is why light doesn't interact with light,
  as we learned in our introductory class on electromagnetism. Actually,
  this is correct only at the classical level. At sufficiently high order
  in perturbation theory, $\gamma \gamma \rightarrow \gamma \gamma$
  scattering is possible, and has been detected by the ATLAS
  collaboration \cite{Agg}.}, the target photon has to split into a
nearly collinear quark--antiquark pair first; one of these will then
interact with the probing photon, thereby giving it a large transverse
``kick'', and allowing both the quark and the antiquark to be physical.
(A massless photon cannot transition into two physical massive particles.)

\begin{figure}[t]
   \centering
   \includegraphics*[width=14cm]{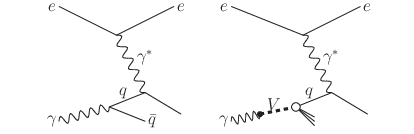}
   \caption{\textit{Deep--inelastic scattering on a real photon
       $\gamma$ leading to a multi--hadron final state can either
       proceed via the basic point--like $\gamma q \bar q$ coupling
       (left diagram), or via the transition of the photon into a slightly
       virtual vector meson $V$ on which the electron then scatters
       (right diagram). The existence of the latter type of contribution
       shows that the parton densities in the photon have a contribution
       that cannot be computed perturbatively.}}
%   \vspace*{-2mm}
   \label{fig3}
\end{figure}

This is depicted in Fig.~\ref{fig3}. While the formal similarity to
the second diagram in Fig.~\ref{fig1} is evident, there is an
important physical difference. No small coupling is necessary to
``produce'' a parton in a proton. After all, at a very basic level a
proton consists of three quarks; we saw above that these can radiate
collinear gluons with very large probability, {\em if} probed at large
momentum transfer.  The PDFs of the proton are therefore ${\cal O}(1)$
quantities. In contrast, the diagrams in Fig.~\ref{fig3} always start
with a $\gamma \rightarrow q \bar q$ splitting, which involves the
(small) electromagnetic coupling constant. Since the PDFs describe
fluxes, which are related to probabilities i.e. to squared transition
amplitudes, the photonic PDFs are ${\cal O}(\alpha_{\rm em})$
quantities, where $\alpha_{\rm em} \simeq 1/137$ is the (original)
fine structure constant.\footnote{Recent fits of the PDFs of the
  proton \cite{pg1, pg2} also include photons as constituent of the
  proton; after all, a quark can emit a nearly collinear photon, just
  as it can emit a gluon, albeit with reduced probability. The photon
  PDF in the proton is also an ${\cal O}(\alpha_{\rm em})$ quantity.}

If this was the end of the story, resolved photon contributions would
need to be treated together with higher--order QCD corrections.  These
are certainly relevant for a precise reproduction of data, but they
rarely change leading order predictions qualitatively. However, we saw
above that increasing $Q$ increases the likelihood for collinear
splittings. The same is true for the $\gamma \rightarrow q \bar q$
splitting which is the origin of the photonic PDFs. The DGLAP
equations imply that the PDFs of the proton do depend on $Q$. However,
their basic normalizations (more exactly, some $x-$integrals over some
combination of PDFs) do not.  In contrast, all photonic PDFs grow
logarithmically with $Q$. As has first been pointed out by Witten in
1977 \cite{Witten}, this means that photonic PDFs should be considered
to be ${\cal O}(\alpha_{\rm em} / \alpha_S)$ quantities, where
$\alpha_S$ is the strong coupling (the QCD analogue of the fine
structure constant). Note that due to asymptotic freedom discussed
above, $\alpha_S$ decreases logarithmically with increasing $Q$.

I have emphasized that all photonic PDFs originate from a
$\gamma \rightarrow q \bar q$ splitting. Since this involves three
pointlike particles, one might be tempted to conclude that therefore
the photonic PDFs should be computable perturbatively. Indeed, in some
sense this is correct, at least for asymptotically large momentum
transfer \cite{Witten, BB}. Unfortunately at next--to--leading order
this ``asymptotic'' prediction leads to photonic PDFs becoming {\em
  negative} at small scaled energy $x$ when applied at finite $Q$.
Note also that in the same ``asymptotic'' limit all PDFs for the
proton collapse to (properly normalized) $\delta-$functions at $x=0$;
this is evidently very far from the truth even for the TeV--scale
momentum transfer probed in $pp$ scattering at the Large Hadron
Collider at CERN.

The correct procedure \cite{Rossi, GGR} is therefore very similar to
that used for determinations of the PDFs of the proton: one starts
from some input at a relatively low value of $Q$, and uses modified
DGLAP equations \cite{apgam} to predict the PDFs at higher $Q$; the
input distributions are fitted from data. The DGLAP equations need to
be modified since the $\gamma \rightarrow q \bar q$ splitting is
treated on the same footing as, e.g., the splitting of a gluon into a
$q \bar q$ pair.

The existence of a not perturbatively computable contribution to the
photonic PDFs can also be motivated from the argument that a photon
can transition to a (virtual) neutral vector meson, like the
$\rho^0\,, \omega$ or $\phi$, see the right diagram in Fig.~\ref{fig3}.
This ansatz has e.g. been used with some success to describe nucleon
form factors \cite{VMD}. The PDFs in these vector mesons are as
non--perturbative as those in the proton.

The upshot of the discussion so far is that at sufficiently large
energies, above several GeV, one can assign parton, i.e. quark and
gluon, densities to the photon. These differ from the parton densities
in the proton by being quantities of order $\alpha_{\rm em} / \alpha_S$,
rather than being ${\cal O}(1)$ quantities. In the next two Sections we
will see what this means for interactions of very energetic photons
with protons and with other photons.

\section{Resolved Photons at $ep$ Colliders}

At the beginning of the previous Section, $ep$ scattering via the
exchange of a photon with virtuality $Q$ was discussed. The
corresponding cross section is suppressed by $1/Q^2$. Reactions with
small $Q$ are therefore much more likely. Kinematically the lower
bound on $Q$ is determined by the electron mass,
$Q_{\rm min} \simeq m_e$. This value can be neglected relative to
typical hadronic scales, which are set by the mass of the pion or
proton. It has been recognized more than 90 years ago \cite{WW} that,
to good approximation, one can therefore often treat the photons
emitted by very energetic electrons as being quasi--real. Many $ep$
collisions can therefore be considered as photoproduction reactions
initiated by the emission of an almost on--shell photon off the
electron. To that end, the flux of nearly on--shell photons can be
treated very much like a parton density, although now it's a photon
density in the electron rather than a quark or gluon density in the
proton; this photon density, usually denoted by $f_{\gamma|e}$, is of
order $\alpha_{\rm em} \ln(E_e/m_e)$, where $E_e$ is the energy of the
beam electrons.

Recall, however, that we need a large momentum transfer, or large
virtuality, somewhere in the process for it to be describable by
perturbative QCD. The simplest QCD processes which satisfy this
condition lead the the production of two jets with sizable transverse
momentum (``transverse'' relative to the direction of the incoming
beams). Since the photon is essentially collinear to the electron, its
transverse momentum is very small; the two jets therefore have equal
and opposite transverse momenta. Perturbative QCD is applicable if the
corresponding absolute value $p_T$ is larger than a few GeV.

\begin{figure}[t]
   \centering
   \includegraphics*[width=14cm]{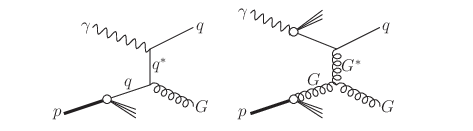}
   \caption{\textit{Contributions to dijet production in photon proton
       scattering; in the examples shown, one quark and one gluon is
       produced at large transverse momentum. The direct process
       (left) proceeds via the exchange of a virtual quark $q^*$ in
       the $t-$channel; there is also a contribution where $q^*$ is
       exchanged in the $s-$channel. The resolved photon process
       (right) proceeds via the exchange of a virtual gluon $G^*$;
       again a contribution from $q^*$ exchange in the $s-$channel
       also exists. Direct and resolved photon contributions have to
       be added incoherently, since only the latter include photon
       remnants in the final state. In addition there are
       contributions with a $q \bar q$ pair in the final state;
       resolved photon processes can also lead to two quarks, two
       antiquarks or two gluons in the final state \cite{DG1}.}}
%   \vspace*{-2mm}
   \label{fig4}
\end{figure}

Rohini and I predicted \cite{DG1} cross sections of such reactions for
the $ep$ collider HERA operating at an $ep$ center--of--mass energy
$\sqrt{s} = 314$ GeV, a few years before it started operations at DESY
in Hamburg. The discussion of the previous Section shows that to
leading order in the strong coupling $\alpha_S$ one not only has to
consider ``direct'' contributions, where the entire energy of the
photon goes into the hard scattering, but also ``resolved''
contributions, where instead a parton from the photon initiates the
hard scattering; see Fig.~\ref{fig4}. Indeed, we found that the latter
{\em dominate} for $p_T \leq 30$ GeV.

In order to understand this, note first of all that both kinds of
photoproduction reactions occur at the same order in QED and QCD
coupling constants. Direct contributions involve one QED and one QCD
vertex in the hard scattering process; an example is shown in the left
diagram of Fig.~\ref{fig4}; the cross section is this
${\cal O}(\alpha_{\rm em} \alpha_S)$. Resolved photon contributions
involve two QCD vertices, so the hard, partonic cross section is
${\cal O}(\alpha_S^2)$. Since the parton density in the photon is
${\cal O}(\alpha_{\rm em}/\alpha_S)$, the resulting contribution is
therefore also ${\cal O}(\alpha_{\rm em} \alpha_S)$. This had not been
appreciated previously. Moreover, only resolved photon processes
receive contributions where a gluon is exchanged in the $t-$ or
$u-$channel; an example is shown in the right diagram of
Fig.~\ref{fig4}. This, as well as color factors, enhances the cross
section. On the other hand, at large $p_T$ the direct contributions
win since only there the entire energy of the incoming photon goes
into the hard scattering, as already noted.

This last point also leads to a qualitative difference between the
final state produced by these two classes of contributions. It always
contains at least two jets with large $p_T$ (possibly more once higher
order contributions are included), as well as the remnants of the
proton; the latter go mostly into the direction of the incoming
proton.  Resolved photon processes {\em in addition} contain the
remnants of the photon from which a parton is ``pulled'' to
participate in the hard scattering; this second set of remnants goes
mostly in the direction of the incoming electron, i.e. opposite to the
direction of the proton remnants. Detectors at high--energy particle
colliders can properly detect and measure only particles that have an
angle of at least a few degrees relative to the beam
directions. Nevertheless a resolved photon event will typically
contain a few (usually not very energetic) hadrons detected not far
from the direction of the outgoing electron.  This difference in final
states also implies that the two contributions have to be added {\em
  incoherently}. However, the properties of these remnants cannot be
predicted from first principles in perturbative QCD.

Our predictions, backed up by the computation of next--to--leading
order (NLO) corrections by several groups \cite{NLO1}, where confirmed
by experiment. For example, in \cite{zeus} the ZEUS experiment reported
that even after requiring $p_T > 14$ GeV, some $70\%$ of the events
were due to resolved photon processes.

The HERA collider ceased operations in 2007. Currently the ``electron
ion collider'' is being constructed in Brookhaven in the US. In the
$ep$ mode it will have a center--of--mass energy $\sqrt{s} \simeq 100$
GeV; an upgrade may reach $\sqrt{s}=140$ GeV. These energies would
definitely allow to probe resolved photon processes; this apparently
has so far ``only'' been studied in conference proceedings \cite{GuKla}.
Furthermore, there is a plan for another electron ion collider in
China \cite{eicc}. However, its center of mass energy would not exceed
$20$ GeV, which is probably too small to probe resolved photons in the
perturbative regime.

\section{Resolved Photons at $e^+e^-$ Colliders}

$e^+e^-$ colliders have been, and hopefully will be, built mostly in
order to study $e^+e^-$ annihilation events. Here the entire energy of
both colliding particles goes into the final state. This maximizes the
reach in the mass of heavy particles that can be produced. Moreover,
it implies complete knowledge of the $4-$momentum of the final state,
which greatly facilitates kinematic reconstruction. Finally, there is
no ``underlying'' event due to (e.g.) proton remnants, which always
occur in hard reactions involving protons. Experiments at $e^+e^-$
colliders therefore operate in a much cleaner environment than those
at $ep$ or $pp$ colliders.\footnote{These arguments ignore initial
  state radiation (ISR), i.e. the emission of nearly collinear photons
  from the initial state.  Due to the small size of $\alpha_{\rm em}$
  the probability for such emission is large only if the photon is
  nearly collinear and fairly ``soft'', i.e.  has a rather small
  energy. The kinematics of typical ISR events is therefore not very
  different from those without ISR. Moreover, ISR emission can be
  described perturbatively in QED, and can thus be corrected for. In
  contrast, the proton remnants typically carry a large but unknown
  fraction of the energy of the incoming proton, and cannot be
  described perturbatively.}

Recall, however, that each electron (or positron) comes with its own
flux $f_{\gamma|e}$ of quasi--real photons. An $e^+e^-$ collider is therefore
also an $e\gamma$ and a $\gamma \gamma$ collider. In fact, deep inelastic
scattering on (quasi--)real photons, discussed in Sec.~1, has so far
only been performed at $e^+e^-$ colliders \cite{f2gam}. Here we are concerned
with the collision of two quasi--real photons.

At first sight one might think that such reactions occur much less
frequently than $e^+e^-$ annihilation reactions. After all, cross
sections for the latter are quadratic in the fine structure constant,
while cross sections for the former involve four powers of
$\alpha_{\rm em}$, two from the cross section for producing the final
state from $\gamma \gamma$ collisions and two from the two flux
factors $f_{\gamma|e}$. However, the flux factors also bring in two
factors of $\ln(E_e/m_e)$, which amounts to about $150$ for
$E_e = 100$ GeV as at the final stages of LEP. More importantly, the
annihilation cross section is suppressed by the square of the $e^+e^-$
center of mass energy $s$, while the cross section for two--photon
collisions scales with the inverse of the squared center of mass
energy $\hat s$ of the innermost reaction; for example, for the
production of a muon pair in two--photon collisions this can be of the
order of $m_\mu^2$. As a result, already at $\sqrt{s} \simeq 5$ GeV the
main source of $\mu^+\mu^-$ pairs produced at an $e^+e^-$ collider is
$\gamma\gamma$ collisions, not $e^+e^-$ annihilations.  Similarly, for
$\sqrt{s}$ exceeding some tens of GeV, the main source of dijet events
will be $\gamma\gamma$ collisions, not $e^+e^-$ annihilation. Of
course, the former will tend to produce much less energetic
(``softer'') jets than the latter. Nevertheless jet production in
$\gamma\gamma$ collision is not only of interest in its own right, it
is also an important background for certain $e^+e^-$ annihilation
events, e.g. in searches for invisible particles which carry away a
large fraction of the $e^+e^-$ energy leaving only a relatively small
detectable energy.

\begin{figure}[t]
   \centering
   \includegraphics*[width=14cm]{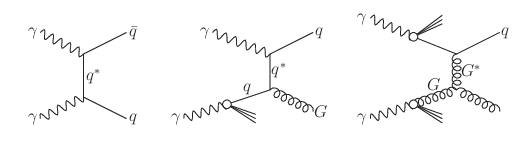}
   \caption{\textit{Contributions to dijet production in photon photon
       scattering. There are now three distinct classes of
       contributing processes: direct (left), single resolved (center)
       and double or twice resolved (right). Contributions where a
       gluon is exchanged in the $t-$channel only exist if both
       photons are resolved. The three classes of processes have zero
       (left), one (center) or two (right) hadronic remnants from the
       resolved photons; they also predict somewhat different angular,
       or rapidity, distributions for the produced jets \cite{DG2}.}}
%   \vspace*{-2mm}
   \label{fig5}
\end{figure}

Rohini and I studied such reactions in \cite{DG2}. There are now three
classes of contributions, since zero, one or two photons can be
resolved, i.e. participate via their partonic constituents, see
Fig.~\ref{fig5}. All three classes contribute at
${\cal O}(\alpha_{\rm em}^2)$ to the $\gamma\gamma$ cross section,
hence at ${\cal O}[\alpha_{\rm em}^4 \ln^2(\sqrt{s}/m_e)]$ to the
$e^+e^-$ scattering cross section. We predicted that for experiments
at the Japanese TRISTAN collider operating at $\sqrt{s} \simeq 60$
GeV, resolved photon contributions should dominate dijet production
for $p_T \leq 3$ GeV at least. In fact, the identification of such
``minijets'', and their {\em quantitative} description \cite{tristan},
which required the inclusion of resolved photon processes, may well be
the main achievement of the TRISTAN program.\footnote{TRISTAN
  experiments did not find any new particles; when these experiments
  were planned it was hoped that the top quark could be discovered
  there. Electroweak and QCD measurements using annihilation events at
  TRISTAN, which operates near the minimum (!) of the annihilation
  cross section before it begins to rise again due to the $Z^0$
  resonance, were soon superseded by measurements taken at LEP and
  SLC.} Among other things it lends credence to the idea that the
rapidly rising inclusive cross section for the production of such
minijets drives the increase of the total inelastic cross sections for
the scattering of hadrons; this not only includes $pp$ and $p \bar p$
scattering, where the cross section has been measured over several
orders of magnitude of center of mass energy, but also $\gamma p$ and
even $\gamma \gamma$ scattering. This is discussed in much more detail
in the contribution by Pancheri and Srivastava to this memorial
volume.

At TRISTAN the production of jets in $\gamma\gamma$ collisions was
only measurable for $p_T \leq 8$ GeV or so. The TRISTAN results were
confirmed by measurements taken at the LEP collider \cite{lep1} at
CERN, using early data taken near the $Z^0$ pole, $\sqrt{s} \simeq 91$
GeV; due to the limited statistics these measurements didn't really
improve on the TRISTAN results. For some reason no analysis of jet
production from $\gamma\gamma$ collisions were published that use
the high integrated luminosity taken at $\sqrt{s} \geq 200$ GeV; these
data could have extended the coverage to $p_T \simeq 20$ GeV or so.

Since at yet higher center of mass energies the $e^+e^-$ annihilation
cross section will keep falling (unless there is a new ``$Z'$''
resonance) while the cross section for $\gamma\gamma$ reactions with a
fixed ``hardness'' $\sqrt{\hat{s}}$ will keep increasing, two--photon
reactions will become relatively ever more important at higher
energies \cite{DG3}. Moreover, in order to keep the number of
annihilation events at least roughly constant, the luminosity will
have to increase $\propto s$. By far the most economical way to
achieve this is to reduce the transverse size of the colliding bunches
(just) before the interaction point. This also increases the strength
of the electromagnetic field caused by the bunch, which in turn
accelerates the particles in the colliding bunch just before (and
after) the collision. Since accelerated electrons and positrons
radiate photons, this ``beamstrahlung'' \cite{beam} can greatly
enhance the flux of (in this case actually) real photons at the
interaction point. As a result, we found \cite{DG4} that some designs
for linear $e^+e^-$ colliders operating at $\sqrt{s} = 500$ GeV that
were discussed in the early 1990's would have had more than one
hadronic $\gamma\gamma$ collision at each bunch crossing. The energy
deposited by these hadrons would have been much smaller than that
associated with ``overlapping events'' at the most recent LHC run, let
alone the upcoming runs where ${\cal O}(50)$ $pp$ collisions are
expected to occur at each bunch crossing. Nevertheless this does cast
some doubt on the claim that $e^+e^-$ colliders always provide a
``very clean environment''.

Unfortunately not much attention is being paid to these effects in the
discussion of the physics potential of future $e^+e^-$ colliders. For
example, according to \cite{fcc} the safest method for measuring the
mass and decay width of the $W$ boson with high precision at a future
(circular) $e^+e^-$ collider is to measure the $W^+W^-$ production
cross section (including contributions from slightly off--shell $W$
bosons) at $\sqrt{s} = 157.5$ and $162.5$ GeV.  In order to fully
exploit the expected statistical accuracy of this measurement, ``the
point--to--point variation of the detector acceptance \dots must be
controlled within a few $10^{-4}$''. Ignoring beamstrahlung, the cross
section for multi--hadron production from $\gamma\gamma$ collisions
will increase by several percent between these two values of
$\sqrt{s}$ \cite{DG5}, the exact increase being quite uncertain. The
effect of this background on the proposed measurement of the mass of
the $W$ boson is currently unclear.

In the even more distant future, a multi--TeV $\mu^+\mu^-$ collider
\cite{muon} may be built. The rate of two--photon reactions at such a
collider would be somewhat smaller than that at an $e^+e^-$ collider
operating at the same energy, due to the $\ln(\sqrt{s}/m_\ell)$
scaling of the photon flux function, where $m_\ell$ is the mass of the
charged lepton; however, at $\sqrt{s} = 10$ TeV this leads to a
reduction of the rate of $\gamma\gamma$ reactions by only a factor of
$2$. While it has been recognized that very energetic muons always
come with a flux of photons (and their hadronic constituents) as well
well as electroweak gauge bosons (and their constituents) \cite{Han},
the implications of this hadronic nature of muon beams for various
backgrounds remain to be explored. Of course, experiments at a muon
collider would have to cope with other potentially very serious
backgrounds as well, caused by the decay products of beam muons.

\section{Summary}

Rohini Godbole made several contributions to the theory of strong
interactions. Among those she was especially proud of her work on
resolved photons, with good reason. We showed that these processes
frequently dominate the hard, and hence perturbatively treatable,
production of multi--hadron (jet) events both in photoproduction
($\gamma p$ scattering) and in $e^+e^-$ collisions. The consequences
of this for hadronic backgrounds at future high energy $e^+e^-$ and
$\mu^+\mu^-$ colliders remain poorly understood.

\subsection*{Reminiscences}

\begin{figure}[h!]
   \centering
   \includegraphics*[width=10cm]{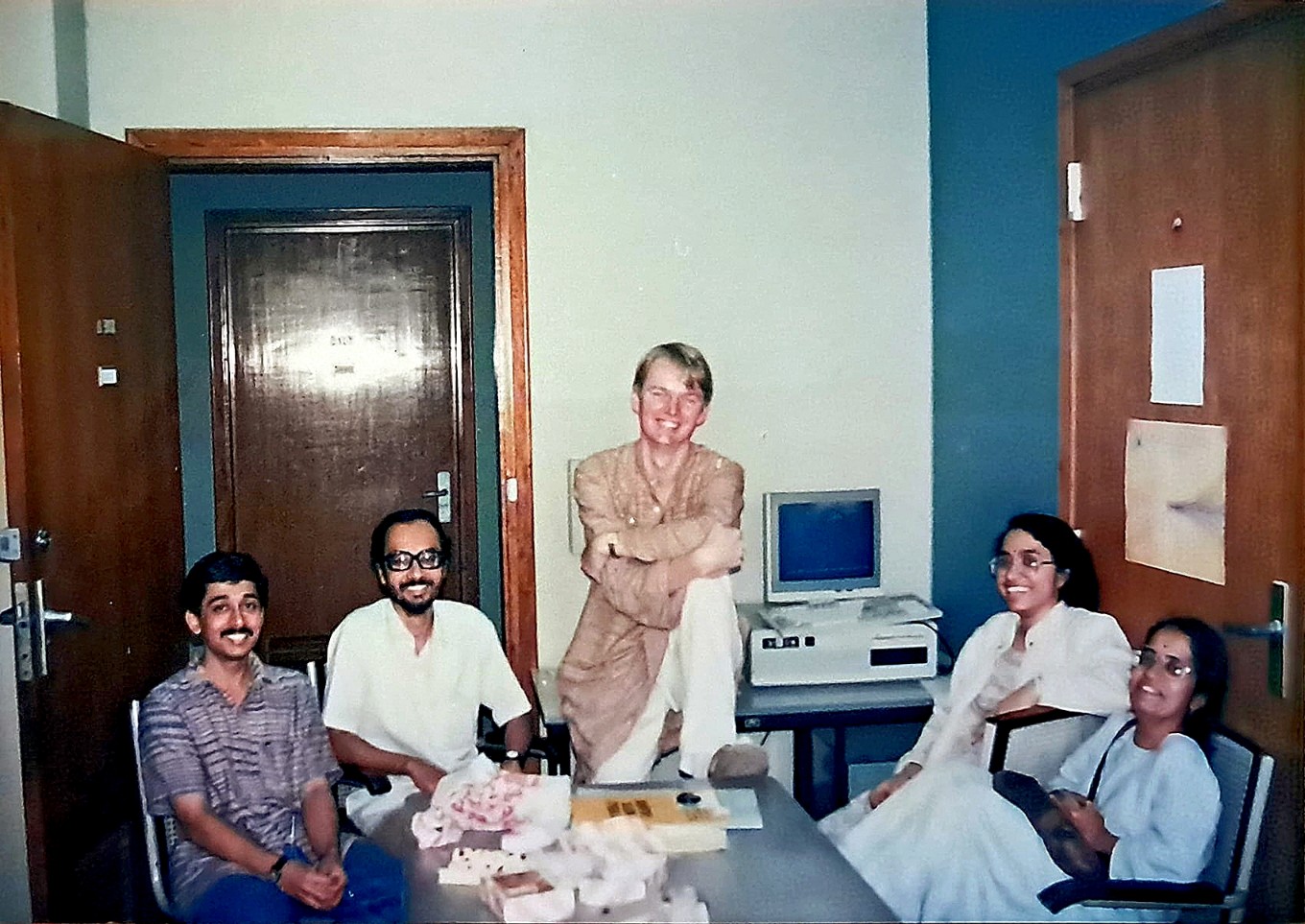}
   \caption{\textit{Rohini (right) and the author (center) at
       TIFR, Mumbai, in the winter of 1992/3. The other people in
       the photo are, from right to left, Sumathi Rao, Ashoke Sen and
     Dileep Jatkar; the photo was taken by Debajyoti Choudhury.}}
%   \vspace*{-2mm}
   \label{fig6}
\end{figure}

In 1992/3 I spent about two and a half months at the TIFR in Mumbai
(then still called Bombay). Rohini was then a professor at Bombay
University, and we spent this time essentially completing our first
review article on resolved photons \cite{DG6}.

Since Mumbai is quite warm (but not unpleasantly hot) even in winter,
I usually wore shorts and a shirt--sleeved shirt. At some point Rohini
decided that I should try a light Indian outfit, and took me shopping.
The result can be seen in Fig.~\ref{fig6}. The photo was taken by
Debajyoti Choudhury, who then was a postdoc at TIFR (and later become
professor at Delhi University). Other then Rohini and myself, it shows
Dileep Jatkar, who also was a postdoc at TIFR at the time; Ashoke Sen,
then a staff member at TIFR; and Sumathi Rao, who was visiting from
the Institute of Physics in Bhubaneswar. Sumathi, Ashoke and Dileep
all later joined the HRI near Prayagraj (Allahabad). There are some
Indian sweets on the table, which we all liked. Unfortunately my
Indian clothes did not survive contact with a US washing machine.

%\begin{thebibliography}{99}   % Use for  10-99  references

\end{document}